\documentclass[aps,prl,twocolumn,superscriptaddress,amsmath,amssymb,showpacs,nofootinbib]{revtex4}

\usepackage{graphicx}
\usepackage{dcolumn}
\usepackage{bm}
\usepackage{color}

\newcommand{\beq}{\begin{eqnarray}}
\newcommand{\eeq}{\end{eqnarray}}

\makeatletter
\AtBeginDocument{\@ifpackageloaded{natbib}{\ifNAT@numbers\if@filesw\immediate\write\@auxout{\string\global\string\NAT@numberstrue}\fi\fi}{}}
\makeatother
\begin{document}

\title{
$^3P_2$ Superfluids Are Topological
}

\author{Takeshi Mizushima}
\affiliation{Department of Materials Engineering Science, Osaka University, Toyonaka, Osaka 560-8531, Japan}
\author{Kota Masuda}
\affiliation{Department of Physics, The University of Tokyo, Tokyo 113-0033, Japan}
\affiliation{Theoretical Research Division, Nishina Center, RIKEN, Wako 351-0198, Japan}
\author{Muneto Nitta}
\affiliation{Department of Physics, and Research and Education Center for Natural Sciences,
Keio University, Hiyoshi 4-1-1, Yokohama, Kanagawa 223-8521, Japan}

\date{\today}

\begin{abstract}
We clarify topology of $^3P_2$ superfluids which are expected to be realized in the inner cores of neutron stars and cubic odd-parity superconductors. $^3P_2$ phases include uniaxial/biaxial nematic phases and nonunitary ferromagnetic and cyclic phases. We here show that all the phases are accompanied by different types of topologically protected gapless fermions: Surface Majorana fermions in nematic phases and a quartet of (single) itinerant Majorana fermions in the cyclic (ferromagnetic) phase. Using the superfluid Fermi liquid theory, we also demonstrate that dihedral-two and -four biaxial nematic phases are thermodynamically favored in the weak coupling limit under a magnetic field. It is shown that the tricritical point exists on the phase boundary between these two phases and may be realized in the core of realistic magnetars. We unveil the intertwining of symmetry and topology behind mass acquisition of surface Majorana fermions in nematic phases.
\end{abstract}


\maketitle

{\it Introduction.}--- 
The topological concept of matter has recently spread over diverse fields in condensed matter. Nontrivial topology embedded in bulk brings about topological quantization in transport and anomalous electromagnetic responses~\cite{qiRMP11,andoJPSJ13}. The topological viewpoint has also shed light on a new facet of unconventional superconductors/superfluids~\cite{volovik,mizushimaJPSJ16,mizushimaJPCM15,tanakaJPSJ12,satoJPSJ16,SchnyderPRB08,satoPRB10,schnyderJPCM}. The key ingredient is Majorana fermions, which behave as non-abelian anyons~\cite{read,ivanov} and possess Ising spins~\cite{chungPRL09,nagatoJPSJ09,volovikJETP09v2,mizushimaPRL12}. The former is expected to be a key for realizing fault-tolerant quantum computation~\cite{Kitaev,nayakRMP08}, while the latter is a consequence of intertwining of topology and symmetry~\cite{mizushimaPRL12}. 




The purpose of this Letter is to unveil topological superfluidity relevant for high dense cores of neutron stars and cubic superconductors~\cite{volovikJETP85,uedaPRB85,ozaki,sigrist}. Neutron stars are unique astrophysical objects under extreme conditions. Neutron superfluidity is an indispensable ingredient for understanding the evolution of neutron stars. Superfluid transitions reconstruct low energy structures of neutrons and considerably affects evolution and cooling of neutron stars. Superfluidity indeed gives a key to understand long relaxation time observed in the sudden speed-up events of neutron stars~\cite{baym69,pines,takatsukaPTP88}, and enhancement of neutrino emmission at the onset of superfluid transition might explain the the recently observed cooling process~\cite{heinke10,pagePRL11,yakovlev,page}. The existence of superfluid components may also explain sudden changes of spin periods observed in pulsars~\cite{reichley,anderson75}. 

\begin{figure}[t!]
\includegraphics[width=80mm]{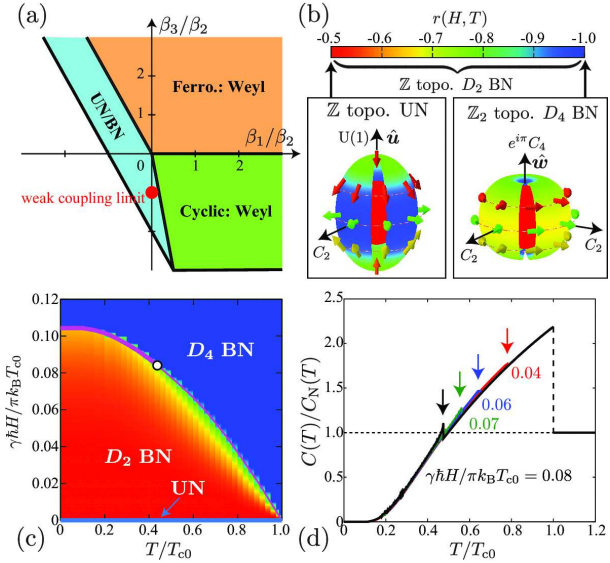}
\caption{(Color online) (a) GL Phase diagram. (b) Gap and topological structures of nematic phases. The thick arrows represent the ${\bm d}$-vectors and the inner (red-colored) sphere denotes the Fermi sphere. (c) Phase diagram under a magnetic field, obtained from the superfluid Fermi liquid theory. The UN phase is stabilized at $H=0$ for $T<T_{\rm c0}$. The thick (thin) curve is the first (second) order phase boundary. (d) $C(T)/C_{\rm N}(T)$ under fixed magnetic fields.}
\label{fig:phase}
\end{figure}

The prediction of $^3P_2$ superfluidity in neutron stars dates back to 1970~\cite{tamagakiPTP70,hoffberg}. A strong spin-orbit force between nuclei generates a short-ranged attractive $^3P_2$ interaction, and the high density induces a repulsive core in the $^1S_0$ channel. The $^1S_0\!\rightarrow\!^3P_2$ transition indeed occurs at the critical density ($\sim\! 10^{14}{\rm g}/{\rm cm}^{3}$) relevant for the interior of neutron stars~\cite{tamagakiPTP70,hoffberg,takatsukaPTP71,takatsukaPTP72,fujitaPTP72,richardsonPRD72}. As seen in Fig.~\ref{fig:phase}(a), superfluid states subject to the total angular momentum $J\!=\! 2$ are classified into several phases~\cite{merminPRA74,saulsPRD78,richardsonPRD72}. Nematic phases preserve the time reversal symmetry (TRS), while the cyclic and ferromagnetic phases are non-unitary states with spontaneously broken TRS. The richness of $^3P_2$ order parameters brings about various types of massive/massless bosonic modes~\cite{bedaquePRC03,leinsonPLB11,leinsonPRC12,leinsonPRC13,bedaquePRC13,bedaquePRC14E,bedaquePRC14,bedaquePLB14,bedaquePRC15} and exotic topological defects, including spontaneously magnetized vortices, fractional, and non-abelian vortices~\cite{muzikarPRD80,saulsPRD82,fujitaPTP72,masudaPRC16,masuda16}. In contrast to ``bosonic'' excitations, there have been no studies on the topology of ``fermions'' in $^3P_2$ superfluids. 

In this Letter, we clarify that various types of topological fermions exist in $^3P_2$ superfluids. Low-lying fermionic excitations in nematic phases are governed by two-dimensional Majorana fermions bound to surface. Their mass acquisition is prohibited by the intertwining of symmetry and topology. In contrast, the cyclic phase possesses eight Weyl points and the low-lying quasiparticles behave as a quartet of itinerant Majorana fermions. These observations on topological fermions may give a new insight into transports and cooling mechanism in the inner cores of neutron stars.

$^3P_2$ phases can be realized in cubic odd-parity superconductors, {\it i.e.}, the $E_u$ irreducible representation of the $O_h$ symmetry group~\cite{volovikJETP85,uedaPRB85,ozaki,sigrist}. The formation of higher partial wave pairs, {\it e.g.}, $^3P_J$, has also been discussed in cold atoms~\cite{wu1,wu2}. We here argue tangible systems to realize topological phenomena inherent to $^3P_2$ phases.


{\it Phase diagrams.}---
We start to clarify the gap structure and the thermodynamic stability of $^3P_2$ superfluids. We define Pauli matrices, ${\bm \sigma}$ (${\bm \tau}$), in the spin (Nambu) space. The bulk states are determined by the Bogoliubov-de Gennes (BdG) Hamiltonian, $\mathcal{H}\!=\!\frac{1}{2}\sum _{\bm k}{\bm c}^{\dag}({\bm k})\mathcal{H}({\bm k}){\bm c}({\bm k})$, 
\beq
\mathcal{H}({\bm k}) = \left( 
\begin{array}{cc}
\varepsilon({\bm k}) & i{\bm \sigma}\cdot{\bm d}({\bm k})\sigma _2 \\
i\sigma _2 {\bm \sigma}\cdot{\bm d}^{\ast}(-{\bm k}) & - \varepsilon^{\rm T}(-{\bm k}) 
\end{array}
\right),
\label{eq:H}
\eeq
where ${\bm c}^{\dag}({\bm k})\!=\![c^{\dag}_{\uparrow}({\bm k}),c^{\dag}_{\downarrow}({\bm k}),c_{\uparrow}(-{\bm k}),c_{\downarrow}(-{\bm k})]$ denotes the creation and annihilation operators of neutrons in the Nambu space. Here, $\varepsilon({\bm k})$ is composed of the $2\!\times\!2$ single-particle energy subject to the simultaneous rotation of spin and orbital spaces, ${\rm SO}(3)_{\bm J}$, and the Zeeman field $-\gamma \hbar {\bm H}\cdot{\bm \sigma}/2$. Spin-triplet pairs are generally represented by ${\bm d}({\bm k})$ and $^3P_2$ order parameter is given by the second-rank, traceless and symmetric tensor, $d_{\mu i}$, where $d_{\mu}({\bm k})=d_{\mu i}\hat{k}_i$ and $\hat{\bm k} \!=\! {\bm k}/k_{\rm F}$. The repeated indices imply the sum over $(1,2,3)$ or $(x,y,z)$. The quasiparticle excitation energy at zero fields is given by diagonalizing Eq.~\eqref{eq:H} as
$E_{\pm}({\bm k}) \!=\! \sqrt{\varepsilon^2_0({\bm k}) + |{\bm d}({\bm k})|^2 \pm |{\bm d}({\bm k})\times {\bm d}^{\ast}({\bm k})|}$,
where $\varepsilon _0 ({\bm k})\!=\! \frac{1}{2}{\rm tr}\varepsilon({\bm k})$. The Hamiltonian holds the particle-hole symmetry (PHS), $\mathcal{C}\mathcal{H}({\bm k})\mathcal{C}^{-1}=-\mathcal{H}(-{\bm k})$, with $\mathcal{C}=\tau _1 K$, where $K$ is the complex conjugation operator. In addition, the TRS, $ \mathcal{T}\mathcal{H}({\bm k})\mathcal{T}^{-1}=\mathcal{H}(-{\bm k})$ with $\mathcal{T}=i\sigma _2 K$, is preserved when $d_{\mu i}\!\in\! \mathbb{R}$ and ${\bm H}={\bm 0}$.


The ground state is determined by minimizing the Ginzburg-Landau (GL) energy functional $\mathcal{F}$, which is invariant under ${\rm SO}(3)_J$ and a gauge transformation, ${\rm U}(1)_{\varphi}$. The functional is given as
$\mathcal{F}\!=\!\alpha {\rm tr}[dd^{\ast}]+\beta _1 |{\rm tr}d^2|^2+\beta _2 [{\rm tr}(dd^{\ast})]^2+\beta _3 {\rm tr}[d^2d^{\ast 2}]$~\cite{saulsPRD78}.
Depending on $\beta _i$, there are several phases as in Fig.~\ref{fig:phase}(a). 
The ground state at the weak coupling limit is the nematic phase which is represented by~\cite{richardsonPRD72,saulsPRD78,sauls} 
\beq
d_{\mu i} = \Delta(T,H) \left[\hat{u}_{\mu}\hat{u}_{i}
+ r \hat{v}_{\mu}\hat{v}_{i} - (1+r)\hat{w}_{\mu}\hat{w}_{i}
\right],
\label{eq:dd}
\eeq
with a orthonormal triad $(\hat{\bm u},\hat{\bm v},\hat{\bm w})$. This state corresponds to highly degenerate minima of $\mathcal{F}$ with respect to $r\!\in\! [-1,-1/2]$. At $r\!=\! -1/2$, $d_{\mu i}$ is invariant under $D_{\infty} \!=\! {\rm SO}(2) \!\rtimes\! \mathbb{Z}_2 \!\simeq\! {\rm O}(2)$ ($\rtimes$ is a semi-direct product), which is called the uniaxial nematic (UN) phase. As shown in Fig.~\ref{fig:phase}(b), the full gap with the hedgehog ${\bm d}$-vector is accompanied by the ${\rm U}(1)$ axis along $\hat{\bm w}$ and $C_2$ rotation axes in the $\hat{\bm v}$-$\hat{\bm w}$ plane. The biaxial nematic (BN) phase at $r \!=\! -1$ remains invariant under the dihedral-four $D_4$ symmetry, which has $C_4$ and $C_2$ axes. The intermediate $r$ holds the $D_2$ symmetry with three $C_2$ axes.


In Fig.~\ref{fig:phase}(c), we display the phase diagram under a magnetic field. This is obtained by minimizing the Luttinger-Ward thermodynamic potential, 
$\delta \Omega[g] \!=\! \frac{\mathcal{N}_{\rm F}}{2}\int^1_0d\lambda \langle{\rm Tr}
\mathfrak{S}(\hat{\bm k})[ g_{\lambda}(\hat{\bm k},\omega_n)-\frac{1}{2}g(\hat{\bm k},\omega_n)]\rangle$, where $\langle\cdots\rangle \!=\! k_{\rm B}T\sum _n \int \frac{d\hat{\bm k}}{4\pi} \cdots$ denotes the Fermi surface average and sum over the Matsubara frequency $\omega _n \!=\! (2n+1)\pi k_{\rm B} T/\hbar$ ($n\!\in\!\mathbb{Z}$)~\cite{serene,vorontsovPRB03,mizushimaPRB12}. The propagator $g$, which is a $4\!\times\!4$ matrix in the Nambu space, is obtained from the low-energy part of the Matsubara Green's function, and the higer energy part is renormalized into the Fermi liquid parameters~\cite{serene}. The propagator is governed by the equation
\beq
[i\omega _n - v-\mathfrak{S}\{g\},g(\hat{\bm k},{\bm r};\omega_n)]+i{v}^{\rm F}_{\mu}\partial _{r_{\mu}}g(\hat{\bm k},{\bm r};\omega _n)=0,
\label{eq:qct}
\eeq
which is supplemented by the normalization condition, $g^2\!=\!-\pi^2$ (we set $\hbar\!=\! 1$). 
This is the transport-like equation propagating along the classical trajectory of the Fermi velocity ${\bm v}^{\rm F}$.  $g_{\lambda}$ is obtained by replacing $\mathfrak{S}\!\mapsto\!\lambda\mathfrak{S}$. The Zeeman term, $v\!=\!-\frac{1}{2}\frac{1}{1+F^{\rm a}_0}\gamma \hbar{\bm H}\cdot{\rm diag}({\bm \sigma},-\sigma_2{\bm \sigma}\sigma_2)$, is rescaled by the Fermi liquid parameter $F^{\rm a}_0$.
The theory is reliable in the weak coupling limit, $\Delta /E_{\rm F}\!\sim\! T_{\rm c0}/T_{\rm F}\!\ll\! 1$ ($T_{\rm c0}$ is the transition temperature at $H\!=\!0$), and applicable to whole temperatures beyond the GL regime~\cite{serene,vorontsovPRB03,mizushimaPRB12}.
The Fermi liquid behaviors and strong coupling corrections in dense neutrons were investigated in Refs.~\cite{backman73,backman79,sjoberg,reviewnn,vulovic}. 

The $4\!\times\!4$ self-energy matrix $\mathfrak{S}$ contains informations on both quasiparticles and $^3P_2$ pair potentials. The $^3P_2$ pair potentials, which appear in the off-diagonal submatrix of $\mathfrak{S}$, are determined with the spin-triplet anomalous propagator, ${\bm f}$, through the gap equation, 
$d_{\mu i}({\bm r}) \!=\! \frac{V}{2}[ 
\langle f_{\mu}\hat{k}_i\rangle +\langle f_{i}\hat{k}_{\mu}\rangle] - \frac{V}{3}{\rm Tr}\langle f_{\mu}\hat{k}_i\rangle$, where $V\!<\! 0$ is the coupling constant of $^3P_2$ interaction. The diagonal submatrix of $\mathfrak{S}$, $\nu$, represents the Fermi liquid corrections, $\nu \!=\! \frac{F^{\rm a}_0}{1+F^{\rm a}_0} \langle g_{\mu} \rangle\sigma _{\mu}$, where the diagonal submatrix of $g$ is represented by the $2\!\times \! 2 $ matrix $g_0+g_{\mu}\sigma_{\mu}$. The magnetization density is $M_{\mu}/M_{\rm N} \!=\! 1 + \frac{2}{\gamma\hbar H}\langle g_{\mu}\rangle$, where $M_{\rm N}$ denotes the magnetization in the normal state. Hence, the diagonal self-energy describes an effective exchange interaction to spin polarization density of neutrons.

No stable region of nonunitary states is found in Fig.~\ref{fig:phase}(c). According to Fig.~\ref{fig:phase}(a), however, the weak coupling limit is close to the boundary of the cyclic phase and the cyclic phase is nearly degenerate with the UN/BN phases. Therefore, the ground state in Fig.~\ref{fig:phase}(c) may be replaced by the cyclic phase when strong coupling corrections are taken into account.


In Fig.~\ref{fig:phase}(c), the UN (BN) phase appears at $H\!=\!0$ ($H\!\neq\!0$). The magnetic field gives rise to the pair breaking in the momentum region within ${\bm d}({\bm k})\cdot{\bm H}\!\neq\! 0$. Consequently, the UN and $D_2$ BN phases are always accompanied by the pair breaking because of ${\bm d}({\bm k})\cdot{\bm H}\!\neq\! 0$ for any ${\bm H}$. The most favored configuration of ${\bm d}({\bm k})$ under ${\bm H}$ is ${\bm d}({\bm k})\perp{\bm H}$, which can be realized by only the $D_4$ BN phase with the nodal direction aligned to $\hat{\bm w}\!\parallel\!{\bm H}$. 

Two BN phases are separated by the second- (first-) order phase boundary in the higher (lower) $T$ regime. The phase boundaries meet at the tricritical point, $(T/T_{\rm c0},\gamma \hbar H/\pi k_{\rm B}T_{\rm c0})\!=\!(0.45,0.083)$ for $F^{\rm a}_0\!=\! -0.7$. To capture a consequence of the tricritical point, in Fig.~\ref{fig:phase}(d), we plot the heat capacity, $C(T)\!=\! C_{\rm N}(T)-T\partial^2\delta \Omega/\partial T^2$, where $C_{\rm N}(T) = \frac{2\pi^2}{3}\mathcal{N}_{\rm F}k^2_{\rm B}T$ is the heat capacity of normal neutrons and $\mathcal{N}_{\rm F}$ is the density of states of normal neutrons at the Fermi level. The heat capacity contains critical information on the thermal evolution of neutron stars~\cite{yakovlev}. In Fig.~\ref{fig:phase}(d), $C(T)$ shows the double jumps and the jump at the lower $T$ increases as $(T,H)$ approaches the tricritical point.
In recent years, neutron stars having strong field $H\!=\! 10^{13}$-$10^{15}$~G, {\it i.e.}, magnetars, have been observed~\cite{paczynski,thompson95,thompsonAJ96,melatos}. The magnetic field corresponds to $\gamma \hbar H/\pi k_{\rm B}T_{\rm c0}\!\approx\!0.001$-$0.1$ with $T_{\rm c0}\!=\! 0.2~{\rm MeV}$. This indicates that the tricritical point may be realized in realistic magnetars.

The first-order phase boundary is sensitive to $F^{\rm a}_0$, and the region is enlarged (reduced) by negative (positive) $F^{\rm a}_0$. For $F^{\rm a}_0\!=\! 0$, the tricritical point indeed lowers to $0.15T_{\rm c0}$. This is attributed to the difference of the magnetic response. The $D_2$ BN phase which has the hedgehog ${\bm d}$-vector suppresses the magnetization relative to that in the normal state, $|{\bm M}|\!<\! M_{\rm N}$, while the $D_4$ BN phase with a two-dimensional configuration of ${\bm d}$ shows $|{\bm M}|\!=\! M_{\rm N}$ when $\hat{\bm w}\!\parallel\!{\bm H}$. The effective field that neutrons experience is affected by the spin polarization of neutrons as ${H}^{\rm eff}_{\mu}\!=\! H_{\mu}-\frac{F^{\rm a}_0}{1+F^{\rm a}_0}\frac{M_{\mu}}{M_{\rm N}}$. Hence, ${\bm H}^{\rm eff}$ in the UN and $D_2$ BN phases is always enhanced (screened) by the polarized medium for $-1\!<\!F^{\rm a}_0 \!<\!0$ ($F^{\rm a}_0\!>\! 0$), and the enhancement/screening effect is fed back to the spin polarization of neutrons. In contrast, no polarization effect is realized in the $D_4$ BN phase, where ${\bm M}\!=\! {\bm M}_{\rm N}$.

{\it Majorana fermions in nematic phases.}---
Let us now clarify the topological aspect of nematic phases. The nematic phases which preserves TRS ($\mathcal{T}^2\!=\!-1$) and the PHS ($\mathcal{C}^2\!=\!+1$) are categorized to the class DIII in the topological table~\cite{SchnyderPRB08}. The nontrivial topology is represented by the three-dimensional winding number $w_{\rm 3d}\!=\! 1$, similar to that of $^3$He-B~\cite{mizushimaJPSJ16}.
%
The hallmark is the presence of massless Majorana fermions on surfaces. To clarify this, we first present the bound state solution of the BdG equation, $\mathcal{H}(k_x,k_y,-i\partial _z)\varphi ({\bm r})\!=\!E\varphi ({\bm r})$, where $\varphi ({\bm r})$ denotes the four-component wavefunction in the Nambu space. The surface is set to be normal to $\hat{\bm z}$ and the specular boundary condition is imposed on $\varphi$.

\begin{figure}[t!]
\includegraphics[width=85mm]{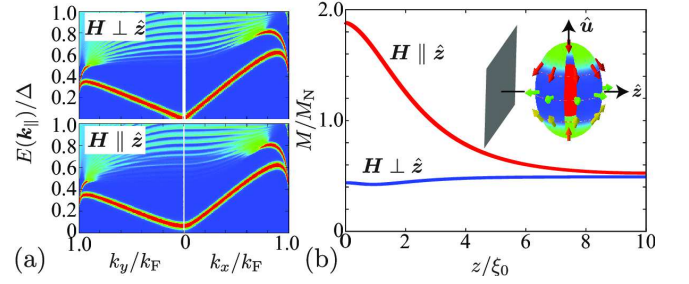}
\caption{(Color online) 
(a) Momentum-resolved surface density of states, $\mathcal{N}_{\rm s}(\hat{\bm k},E)$, and (b) local magnetization for ${\bm H}\perp\hat{\bm z}$ and ${\bm H}\parallel\hat{\bm z}$, in the nematic phase with $\hat{\bm u}\perp\hat{\bm z}$. 
}
\label{fig:nematic}
\end{figure}

In the absence of a time-reversal breaking field, gapless fermions are bound to the surface of nematic phases, which have the relativistic dispersion,
$E(k_x,k_y) \!=\! \pm \sqrt{c^2_xk^2_x + c^2_y k^2_y}$,
for $|E| \!<\! \Delta$. The wavefunction for $E\!>\!0$ is obtained as
$
{\bm \varphi} _E({\bm r}) \!\propto\! 
e^{i(k_xx+k_yy)}\sin(k_{\rm F}\hat{k}_zz)e^{-z/\xi _0}
(
{\bm \Phi}_+
- e^{i\phi _{\bm k}}{\bm \Phi}_-
)$, with the coherence length $\xi_0\!=\!\hbar v^{\rm F}/2\pi k_{\rm B}T_{\rm c0}$ and the spinors 
${\bm \Phi}_+ \!\equiv\! (1,0,0,-i)^{\rm T}$ and ${\bm \Phi}_- \!\equiv\! (0,i,1,0)^{\rm T}$.
The velocities, $(c_x,c_y)$, reflect the orientation of the triad $(\hat{\bm u},\hat{\bm v},\hat{\bm w})$ with respect to the surface:  
$(c_x,c_y) \!=\! \frac{\Delta }{k_{\rm F}}(1,r)$ for $\hat{\bm u}\parallel\hat{\bm z}$, and $\frac{\Delta }{k_{\rm F}}(1,1+r)$ for $\hat{\bm u}\perp\hat{\bm z}$. 

For $T\!\ll\!\Delta/k_{\rm B} $, the field operator can be constructed from only the surface bound states as 
${\bm \Psi}\!=\! (\psi _{\uparrow},\psi_{\downarrow},\psi^{\dag}_{\uparrow},\psi^{\dag}_{\downarrow})^{\rm T}
\!=\! \sum _{E} 
[ {\bm \varphi}_E({\bm r})\eta _{E}
+\mathcal{C}{\bm \varphi}_E({\bm r})\eta^{\dag}_{E}
] $, where $\eta^{\dag}_{E}$ denotes the quasiparticle creation with the energy $0\!<\!E(k_x,k_y)\!<\!\Delta$.
The effective Hamiltonian for gapless surface fermions is given with the spinor $\psi \!=\!(\psi _{\uparrow},\psi _\downarrow)$ and $\bar{\psi}\!=\!i(\sigma _1\psi)^{\rm T}$ as the Majorana Hamiltonian
\beq
\mathcal{H}_{\rm surf} = \int d^2{\bm r}_{\parallel}\bar{\psi}({\bm r}_{\parallel}) ( 
-i\bar{v}_{\mu}\gamma _{\mu}\partial _{\mu} ) \psi ({\bm r}_{\parallel}), 
\label{eq:Hsurf}
\eeq
where $(\bar{v}_1,\bar{v}_2)\!=\!(c_x,c_y)$ and $(\gamma _1,\gamma _2)\!=\! (\sigma _2, -\sigma _1)$.
Hence, the low-energy physics in the nematic phases is governed by Majorana fermions bound to the surface. 

It is remarkable to note that the field operator obeys the Majorana condition
$\psi _{\uparrow} \!=\! i\psi^{\dag}_{\downarrow}$. This
indicates that massless Majorana fermions in Eq.~\eqref{eq:Hsurf} possess the Ising spin character, ${\bm S}\!\equiv\! [\psi^{\dag}_a{\bm \sigma}_{ab} \psi _b - \psi _{a}{\bm \sigma}^{\rm T}_{ab} \psi^{\dag}_b]/4 \!=\!(0,0,S)$. Only perturbation which generates an effective mass in $\mathcal{H}_{\rm surf}$ is an external field coupled to the Ising spin,
$\mathcal{H}_{\rm mass} \!=\! M \int d^2{\bm r}_{\parallel} \bar{\psi}({\bm r}_{\parallel}) ({\bm \sigma}\cdot\hat{\bm z})\psi ({\bm r}_{\parallel})$.
Let us now capture the role of symmetry behind the Ising spin and mass acquisition of surface Majorana fermions. The key is the combined symmetry defined as ${P}_3 \equiv \mathcal{T}C_{2,z}\tau _z$. In the nematic phase, the $C_2$ rotation about $\hat{\bm z}$ denoted by $C_{2,z}\!=\!-i\sigma _z$ only changes ${\bm H}$ to $(-H_x,-H_y,H_z)$. This can be compensated by the TRS ($\mathcal{T}$) and the $\pi$ phase rotation ($\tau _z$) when $\hat{\bm H}\cdot\hat{\bm z} \!=\! 0$. Hence, the operator transforms the BdG Hamiltonian as
${P}_3 \mathcal{H}({\bm k}){P}_3 \!=\! \mathcal{H}(\underline{\bm k})
+\gamma \hbar H {\bm \sigma}\cdot\hat{\bm z}$,
where $\underline{\bm k}\!\equiv\! {\bm k}-2({\bm k}\cdot\hat{\bm z})\hat{\bm z}$ denotes the momentum transfered by ${P}_3$. Only the Zeeman field, $\gamma \hbar{\bm H}\cdot\hat{\bm z}$, breaks the $P_3$ symmetry. For ${\bm H}\cdot\hat{\bm z}\!=\!0$, one can define the chiral operator $\Gamma \!\equiv\! \mathcal{C}{P}_3$ which obeys the chiral symmetry, $\{\Gamma, \mathcal{H}(0,0,k_z)\}\!=\! 0$.
According to the index theorem~\cite{satoPRB11}, one can introduce the winding number along the chiral symmetric momenta ${\bm k}\!=\!(0,0,k_z)$, 
$w_{\rm 1d} \!=\! -\frac{1}{4\pi i}\int dk_z{\rm Tr}
[ 
\Gamma \mathcal{H}^{-1}({\bm k})\partial _{k_z} \mathcal{H}({\bm k})
] _{k_x\!=\!k_y\!=\!0} \!=\! 2$, unless the $P_3$ symmetry is broken. The existence of massless Majorana fermions is guaranteed by $w_{\rm 1d}\!\neq\! 0$.


In Fig.~\ref{fig:nematic}(a), we plot the ${\bm k}$-resolved surface density of states, $\mathcal{N}_{\rm S}(\hat{\bm k},E)\!=\! -\frac{\mathcal{N}_{\rm F}}{\pi}{\rm Im}g_0(\hat{\bm k},0;\omega _n \!\rightarrow\! -iE + 0_{+})$. The surface Majorana fermion acquires mass only when ${\bm H}$ breaks the $P_3$ symmetry. We plot the surface magnetization in Fig.~\ref{fig:nematic}(b). Owing to the Ising character, the gapless surface states do not contribute to the local magnetization density when ${\bm H}\!\perp\!\hat{\bm z}$. In contrast, the massive Majorana fermions in ${\bm H}\!\parallel\!\hat{\bm z}$ significantly enhance the surface magnetization. 
Since neutron stars possess strong axial and toroidal magnetic fields, the Ising spin gives rise to the anisotropic distribution of magnetization on the surface enclosing the $^3P_2$ superfluid core.

\begin{figure}[t]
\includegraphics[width=80mm]{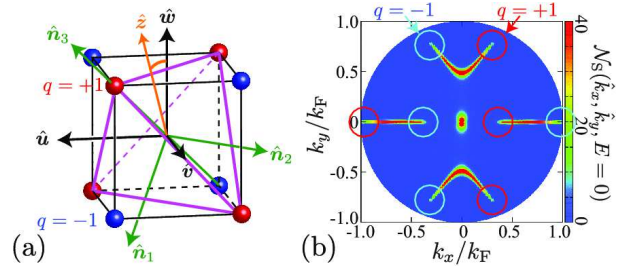}
\caption{(Color online) (a) Configuration of Weyl points in the cyclic phase, where $q\!=\!\pm 1$ possess the monopole charge $q_{\rm m}\!=\! \pm 1$. (b) Momentum-resolved zero-energy density of states on the surface for the misorientation angle $\vartheta \!=\! 2\pi/5$.}
\label{fig:cyclic}
\end{figure}

{\it Cyclic and ferromagnetic phases.}---
The cyclic phase which appears in Fig.~\ref{fig:phase}(a) is the nonunitary state with spontaneously breaking the TRS, whose order parameter is given by replacing $r$ in Eq.~\eqref{eq:dd} to $\omega\!=\!e^{i2\pi/3}$ as 
\beq
d_{\mu i} = \Delta(T,H) \left[\hat{u}_{\mu}\hat{u}_{i}
+ \omega \hat{v}_{\mu}\hat{v}_{i} + \omega^2\hat{w}_{\mu}\hat{w}_{i}
\right],
\label{eq:ddcy}
\eeq
This possesses two distinct gap structures: Full (nodal) gap in the $E_+$ ($E_-$) branch. The nodal points are identified as $q{\bm k}_{\alpha}$ ($\alpha \!=\! 1, \cdots, 4$ and $q\!=\! \pm 1$), where $\alpha$ denotes each vertex of the tetrahedron (Fig.~\ref{fig:cyclic}(a)). The PHS, $\mathcal{C}\mathcal{H}({\bm k}_{\alpha})\mathcal{C}^{-1}\!=\!-\mathcal{H}(-{\bm k}_{\alpha})$, implies that the point node at ${\bm k}_{\alpha}$ must be accompanied by the PHS partner $-{\bm k}_{\alpha}$. 

It is convenient to introduce the new triad $(\hat{\bm n}_1,\hat{\bm n}_2,\hat{\bm n}_3)$, where $\hat{\bm n}_3$ is taken along a nodal direction $q{\bm k}_{\alpha}$ (Fig.~\ref{fig:cyclic}(a)). Let $V_{q,\alpha}$ be a small region around $q{\bm k}_{\alpha}$. In the new basis and the region of ${\bm k}\!\in\!V_{q,\alpha}$, the $4\times 4$ BdG matrix is decomposed into a pair of the $2\times 2$ matrix, $\mathcal{H}({\bm k})\!\mapsto\!{\rm diag}[\mathcal{H}_{+}({\bm k}),\mathcal{H}_{-}({\bm k})]$~\cite{note}, where $\mathcal{H}_{\pm}({\bm k})$ denotes the $E_{\pm}({\bm k})$ branches.
The low-energy effective Hamiltonian in the cyclic phase is therefore governed by the gapless sector, $\mathcal{H}_{-}\!=\! \sum _{q,\alpha}\sum _{{\bm k}\in V_{q,\alpha}}{\bm c}^{\dag}_{\alpha}({\bm k})\mathcal{H}^{q,\alpha}_-({\bm k}){\bm c}_{\alpha}({\bm k})$, which reduces to the Weyl Hamiltonian
\beq
\mathcal{H}^{q,\alpha}_-({\bm k}) = \hat{e}^{\mu}_a\tau^a (k_{\mu}-qk_{\alpha,\mu}),
\eeq
with the vielbein $(\hat{e}^{\mu}_1,\hat{e}^{\mu}_2,\hat{e}^{\mu}_3)\!=\!(\bar{v}\hat{n}_{1,\mu},\bar{v}\hat{n}_{2,\mu},v_{\rm F}\hat{n}_{3,\mu})$, $\bar{v}\!=\!\Delta/k_{\rm F}$, and ${\bm c}_{\alpha}({\bm k})\!=\![c_{\alpha}({\bm k}),c^{\dag}_{\alpha}(-{\bm k})]^{\rm T}$. Each point node is identified as the Weyl point by the monopole charge $q_{\rm m} \!=\! + 1$ ($q_{\rm m}\!=\! -1$) for $q\!=\! +1$ ($-1$), which is a source of the hedgehog-like Berry curvature in ${\bm k}$ space. 
Reflecting the Weyl points, zero energy flat bands appear on the Fermi surface which connect a pair of the Weyl points projected onto the surface. Figure~\ref{fig:cyclic}(b) shows the ${\bm k}$-resolved zero-energy density of states on the surface, $\mathcal{N}_{\rm s}(\hat{k}_x,\hat{k}_y,E\!=\!0)$, where the surface normal axis $\hat{\bm z}$ is assumed to be tilted from $\hat{\bm w}$ by angle $\vartheta\!\equiv\!\cos^{-1}(\hat{\bm z}\cdot\hat{\bm w})$.


We now introduce the coordinate centered on the Weyl point, ${\bm K}_q\!\equiv\! {\bm k}-q{\bm k}_{\alpha}$. The four-component real quantum field, $\psi ({\bm r})\!=\! \mathcal{C}\psi ({\bm r})$, is constructed from a PHS pair of the single-species Weyl fermions as $\psi _{\alpha} ({\bm r})\!\equiv \! \sum _{\bm K}e^{i{\bm K}\cdot{\bm r}}[c_{\alpha}({\bm K}_+), c_{\alpha}({\bm K}_-), c^{\dag}_{\alpha}(-{\bm K}_+), c^{\dag}_{\alpha}(-{\bm K}_-)]^{\rm T}$~\cite{fu16}. The low energy Hamiltonian is governed by massless Majorana fermions
\beq
\mathcal{H} \approx
\mathcal{H}_- = \sum _{\alpha} \int d^3{\bm r} 
\bar{\psi}_{\alpha}({\bm r})
\left[ 
-i\hat{e}^{\mu}_a\gamma ^{a}
\partial _{\mu} \right]
{\psi}_{\alpha}({\bm r}),
\label{eq:3dMF}
\eeq
where we introduced $\bar{\psi}\!=\!(\tau _1 \psi)^{\rm T}$ and ${\bm \gamma} \!=\! (\mu _1\tau _1,\mu _1\tau _2,\mu _3)$ with the Pauli matrices $\mu _{i}$ labeled by the PHS index $q \!=\! \pm 1$. 
The itinerant Majorana fermions with pseudospin $\frac{1}{2}$ form a quartet ($\psi _1,\psi _2, \psi _3,\psi _4$) as a consequence of the tetrahedral symmetry. 

Another phase in Fig.~\ref{fig:phase}(a) is known as the ferromagnetic phase, 
$d_{\mu i} \!=\! \Delta (\hat{u}_{\mu}+i\hat{v}_{\mu}) (\hat{u}_{i}+i\hat{v}_{i}) $.
This state is equivalent to the A$_1$ phase of $^3$He~\cite{vollhardt}. 
Similar to $^3$He-A$_1$, the Zeeman splitting of the Fermi surface in extremely high fields might favor the ferromagnetic phase. The nonunitary phase is accompanied by a single itinerant Majorana fermion with $\uparrow$ spin, {\it i.e.}, $\alpha\!=\! 1$ in Eq.~\eqref{eq:3dMF}.

{\it Cubic superconductors.}---
$^3P_2$ phases can be realized in cubic superconductors as the two-dimensional odd-parity $E_u$ state~\cite{volovikJETP85,ozaki,uedaPRB85,sigrist}. 
The ${\bm d}$-vector is represented by
${\bm d} ({\bm k}) \!=\! \eta _1 {\bm \Gamma}_1({\bm k}) + \eta _2 {\bm \Gamma}_2({\bm k})$,
where the basis functions of the $E_u$ state are given by
${\bm \Gamma}_1({\bm k}) \!=\! (2\hat{\bm c} \hat{k}_c
-\hat{\bm a}\hat{k}_a - \hat{\bm b}\hat{k}_b)/\sqrt{2}$ and
${\bm \Gamma}_2({\bm k}) \!=\! \sqrt{3/2}(\hat{\bm a}\hat{k}_a-\hat{\bm b}\hat{k}_b)$. The 4-th order GL energy relevant to two-component order parameters is given by $\mathcal{F}\!=\! \beta_1(|\eta_1|^2+|\eta_2|^2)^2 + \beta_2(\eta _1 \eta^{\ast}_2-\eta^{\ast}_1\eta_2)^2$. The cyclic order parameter of Eq.~\eqref{eq:ddcy} is obtained as $(\eta _1,\eta _2) \!=\! (1,i)$ for $\beta _2/\beta _1 \!<\! 0$. For $\beta _2 /\beta _1 \!>\! 0$, the UN/BN phases are realized by $(\eta _1, \eta _2 )\!=\! (\cos\theta,\sin\theta)$, where $\theta \!=\! 0$ ($\pi/2$) corresponds to the UN ($D_4$ BN) phase.

{\it Concluding remarks.}---
We have demonstrated that different types of topological fermions exist in $^3P_2$ phases: Surface Majorana fermions in nematic phases and itinerant Majorana fermions in the cyclic and ferromagnetic phases. 
The topological and symmetry protection of neutrons may significantly affect the heat transport and cooling mechanism.
Furthermore, we have mentioned that topological $^3P_2$ phases may be realized in solid states, such as cubic superconductors. The heavy fermion superconductors, PrOs$_4$Sb$_{12}$~\cite{kozii16} and UBe$_{13}$~\cite{PfleidererRMP09}, might be possible candidates to realize $^3P_2$ phases.

The dense core of neutron stars consists of mostly $^3P_2$ superfluid neutrons with a small amount of superconducting protons and normal electrons in beta equilibrium. There are open questions regarding the influence of protons. Firstly, Eq.~\eqref{eq:H} is extended into the form which takes account of a neutron-proton interaction through nuclear forces. When the interaction is weak, the topology of an extended $\mathcal{H}$ is governed by the topology of the majority component, {\it i.e.}, the $^3P_2$ phase. For the general case of the interaction, however, its influence on topology remains as an unresolved problem. Another role of superconducting protons is the expulsion and confinement of magnetic fields. When protons are a type-II superconductor, the magnetic field is confined into a low density of flux filaments within the penetration depth $\Lambda _{\rm p}\!\approx\! 67~{\rm fm}$. Since the mean distance of filaments is much longer than $\Lambda _{\rm p}$, neutrons are free from the field~\cite{baym,sauls_nato}. However, it has been pointed out that the type-II senario is inconsistent with observations of a long period precession in isolated pulsars~\cite{link,buckley,sedrakian}. A type-I superconductor of protons may form the intermediate state with alternating domains of superconducting and normal regions. The inhomogeneous magnetic field leads to the spatially inhomogeneous ground states, since a low (a high) field favors the UN and cyclic ($D_4$ BN) phase. Topologically protected gapless fermions may appear at the interface of domains with different topology. 



We also notice that the richness of $^3P_2$ order parameter manifolds leads to exotic topological excitations, such as non-abelian fractional vortices~\cite{masuda16,semenoff,kobayashiPRL09,kawaguchi}. $^3P_2$ superfluids offer a unique platform to study the interplay between non-abelian Majorna fermions and non-abelian vortices.

This work was supported by JSPS (No.~JP16K05448 and No.~JP25400268) and ``Topological Materials Science'' (No.~JP15H05855) and
``Nuclear Matter in Neutron Stars Investigated by Experiments
and Astronomical Observations'' (No.~JP15H00841) KAKENHI on innovation areas from MEXT. The work of M. N. is also supported in part by the MEXT-Supported Program for the Strategic
Research Foundation at Private Universities ``Topological
Science'' (Grant No.~S1511006).

\bibliography{neutronstar}

\end{document}